\RequirePackage{fix-cm}
\documentclass[smallextended]{svjour3}       
\smartqed  
\usepackage{graphicx}
\begin{document}

\title{Predicting academic success in Belgium and France 
}

\subtitle{Comparison and integration of variables related to student behavior\\}


\author{       
}


\institute{
}

\date{Received: date / Accepted: date}

\maketitle

\begin{abstract}
Having observed low success rates among first-year university students in both Belgium and France, we develop prediction models in this paper in order to identify, at the earliest  possible stage, those students who are at risk of failing at the end of the academic year. We applied different data mining techniques to predict the students' academic success. We find that it is very difficult to predict success by only considering the variables related to behavior during classes, and that it is necessary to add variables related to personal history, involvement in and behavior during their studies, and perceptions of academic life, to obtain good-quality results.

\end{abstract}

\section{Introduction}

In France and Belgium, there have been low success rates among first-year university students for years. About three out of every five students who graduate from high school and enroll in one of the nine French-speaking universities in Belgium fail or drop out~\cite{Vandamme07}. A similar situation is observed in France and in other countries~\cite{Kovacic10}. This alarming report fuels discussions with the different stakeholders in higher education in Belgium's French-speaking Community and with the ministerial supervisory authorities in France, and leads to various actions, financed by grants-in-aid from the state, region, or department or by equity, in order to reduce the substantial economic, social and human costs of such a high failure rate among first-year university students.

The objective of this research is to highlight the factors that explain the academic success of first-year university students, in order to propose an approach to building a model for predicting academic success. With this model, we will be able, at the beginning of the academic year, to classify the students into two categories: those with a high probability of success (HPS) and those with a low probability of success (LPS). To correctly target vulnerable students who really need support measures, we plan to put in place a decision-making tool to identify LPS students as early as possible in the year, i.e. before the first assessments. This will allow us to optimize the distribution of the teaching resources (computer-assisted teaching, tutorials, etc.) to increase academic success. 

This paper is organized as follows. We first give a brief survey of the different studies related to the prediction of academic success. We then present the methodology we adopted (Section~\ref{SecMethodology}). Next, we describe our data (Section~\ref{SecData}) followed by a descriptive analysis (Section~\ref{SecDescriptiveAnalysis}). In section~\ref{SecCorrelations}, we analyze the correlations between the variables and academic success. Finally, in section~\ref{SecMethods}, we present the methods of data mining employed, and in section~\ref{SecResults}, the different results obtained with these methods.

\section{Survey}

\label{SecSurvey}

In the last 20 years, data mining methods have intensively been used in the context of education. Romero and Ventura~\cite{Romero07} have reviewed all the applications of data mining to traditional educational systems, i.e. in enrollment management~\cite{Malt07}, academic performance, graduation, web-based education, etc. 

There are many studies~\cite{Murtaugh99,Noble07,Pascarella83,Simpson06,Strayhorn09,Tharp98} related to the prediction of academic success and we refer to~\cite{Reason09} for a complete overview. It is interesting to note that, depending on the data, the conclusions obtained are rather different. For example, in~\cite{Ishitani03}, Ishitani compared the results obtained by students whose parents are not university graduates (called first-generation students) and students who have at least one parent who graduated from university. He concluded that first-year university students whose parents did not graduate from university have a 71\% higher risk of dropping out than students with two university-educated parents. On the other hand, in~\cite{Pratt89}, Pratt and Skaggs drew the opposite conclusion, i.e. that there is no difference between first-generation students and others in terms of dropout risk.  

Further details of two papers that we found interesting in the context of our study are given below. 

In~\cite{Vandamme07}, Vandamme \emph{et al.} tried to classify students into three groups as early as possible in the academic year: ``low-risk'' students (high probability of succeeding), ``medium-risk'' students and ``high-risk'' students (high probability of failing). A total of 533 Belgian students from different faculties (management science, political science, engineering and bio-engineering) were considered. To identify to which group the students belonged, they used the results obtained by the students during the first examinations in January\footnote{In Belgium, examinations are in January and in June, and for those students who fail, retakes are in September.}. They established the following rule: students who obtained an average mark of less than 45\% have a high risk of failing at the end of the year, and those who obtained an average mark of more than 70\% have a low risk of failing at the end of the year. They used discriminant analysis, neural networks and decision trees to predict the 
category to which the students belonged. Taking into account these three categories, they only obtained an overall total correct classification rate of 57.35\% (with the linear discriminant analysis method), which is substantially worse than if they had been interested in a binary success/failure variable. Therefore, if this ternary approach allows more flexibility than a binary classification, good classification rates are more difficult to achieve. The most important factors identified as correlating to academic success were previous education, the number of hours of mathematics classes, financial independence, and age. 

In~\cite{Kovacic10}, Kovacic studied the influence of socio-demographic variables (age, gender, ethnicity, education, work status, and disability) and study environment (course program and course block) on the academic success of students from the Open Polytechnic of New Zealand. He considered 450 students, between 2006 and 2009, who enrolled in an information systems course. He used different classification trees and obtained an overall classification rate equal to 60.50\%. The most important factors were ethnicity, course program, course block, high school, and early enrollment.
 
It can be seen that, depending on the type of studies, the course, and the university, the results and the factors correlated to academic success can be very different.

\section{Methodology}

\label{SecMethodology}

Based on numerous studies highlighting a multitude of factors influencing academic success, Vandamme \emph{et al.}~\cite{Vandamme07} devised a questionnaire in order to gather the maximum amount of relevant information from first-year university students. We have used the same questionnaire in this study.

The questionnaire is mainly based on an analysis carried out in 1994 by P. Parmentier~\cite{Parmentier94}, who found that the results obtained by students in mid-term or final exams are influenced by three sets of interacting factors. The first set comprises structural or stable factors, while the other two consist of processual or changing factors. 

The first of these sets concerns the personal history of the students, e.g. identity, socio-familial 
past, and educational background. Questions about nationality, when they obtained their high school diploma, type of high school diploma, housing (living with their parents or in student accommodation), their parents' occupation and qualifications, number of siblings, parents' marital status, how they are financing their studies, smoking, drinking, state of health, reasons for choosing this university and this course of study, etc. are all included in the questionnaire. 

The second set of factors can be interpreted as the expression of the students' involvement in his studies and behavior during their studies. Let us mention, for example, participation in optional activities, meeting teachers to ask questions or to request answers to previous exam papers, participation in courses (optional or otherwise), distribution of time between studies and other activities, study method, etc.

The final set of factors relates to the perceptions of the students regarding their studies. The questions are thus related to the way they perceive their teachers, the courses, the university and academic life, the feeling of having chosen the right university, their self-assessment regarding their ability to succeed, etc. 

This questionnaire was distributed to students at Jean Monnet University in Saint-Etienne in France and to students at the Catholic University of Mons in Belgium at the beginning of two academic years: 2009-2010 and 2010-2011. The students had already attended some classes before receiving the questionnaire. 

In 2010-2011, in addition to the standard questionnaire of Vandamme \emph{et al.}~\cite{Vandamme07}, we tried to integrate variables related to student class behavior, which, to our knowledge, are variables that have never been taken into account before. 

To integrate these variables, our initial idea was to analyze student attitudes by sitting in on classes to watch and examine the students. However, this idea was abandoned, as it would have been a time-consuming and arduous exercise. Another option was to install cameras, but aside from the technical difficulties, this is arguably an invasion of privacy. We therefore simply devised a questionnaire containing 14 questions about the behavior of the students during classes. The questions asked will be detailed in Section~\ref{SubsubsecBehaviors} in Table~\ref{StatQuestionsBehaviors}. 

The students also had to indicate their favorite positions in the lecture theater. For this purpose, the lecture theater was represented by a square, divided into nine possible positions (the positions obtained with the combination of (left, center, right) and (front, middle, back)). Students were asked to specify their favorite positions based on these nine possible options, and could indicate more than one favorite. 

We used different data mining techniques (decision tree, random forests, linear discriminant analysis, binary logistic regression, support vector machines and $k$-nearest neighbors) to extract knowledge from this database, enabling us to effectively target those students who most need help. Educational support resources -- necessarily limited -- such as tutoring by an older student or mentoring by a teacher, should be primarily directed to those students with a high risk of failing.

In the next section, we describe the data on which our study is based.

\section{Data}

\label{SecData}

In this paper, we first analyze the results of the survey conducted in 2009-2010 among 614 first-year students from the Institute of Technology at Jean Monnet University in Saint-Etienne, France, and 169 students from the Catholic University of Mons in Belgium.

Of the 614 French students, 88 study management, 75 biology, 52 mechanics, 145 marketing, 79 physics, 119 business administration, and 56 computer science. These students were selected by written application. 

As for Belgium, 37 study communication science, 103 management science, 4 human science, and 25 political science. All these students are graduated from high school, the only entry requirement to the university (contrary to the French students at the Institute of Technology).

The survey was also conducted in 2010-2011 with 52 first-year students from the Faculty of Science of Jean Monnet University and 162 students from the Catholic University of Mons. Of the 52 French students, 6 study applied mathematics and social sciences, and 46 computer science and mathematics. Of the Belgian students, 36 study communication science, 98 management science, 4 human science, and 24 political science. All these students (France and Belgium) graduated from high school, the only entry requirement for the Faculty of Science at Jean Monnet University and the Catholic University of Mons.

Students were asked to complete the questionnaire during a class, which means that the response rate corresponds to the students' attendance rate (in the first year, and particularly at the beginning of the year, the attendance rate was extremely high, being close to 93\%~\cite{Vandamme07}). 

In 2009-2010 the questionnaire, which included a range of questions (almost all closed) covering 42 subjects and incorporated a system guaranteeing student anonymity, led to the creation of a database in which each student is described using certain criteria and attributes. A total of 145 variables were extracted from the questionnaire, such as age, parents' education, and perceptions of academic life. 

In 2010-2011 the questionnaire included 44 subjects (2 subjects concerning student class behavior were added compared with the 2009-2010 questionnaire), and 176 variables were extracted. 

Most of the variables were bounded and discrete, and those that were not were discretized (except the birth year of the student) into a set of four or five values. Some of the variables had missing values, which can be a problem for some data mining methods. We therefore replaced the missing values with probable values, calculated as follows. For a case with unknown values for one of the variables, we identified the 10 most similar cases. The missing value was then replaced with the median for these 10 most similar cases. To define the notion of similarity, the Euclidean distance was used, with normalization of the values of the variables (we normalized the variables such that every variable has a zero mean and a unit standard deviation).

Depending on the students' final results (determined in July in France, and September in Belgium), we allocated to each student a binary variable of success, equal to 1 if the student passed, and 0 otherwise.

\section{Descriptive analysis}

\label{SecDescriptiveAnalysis}

We first give descriptive statistics for the 2009-2010 questionnaire by comparing the students from France and Belgium. We then do the same for the 2010-2011 questionnaire, except that we also study the variables related to student class behavior. 

\subsection{2009-2010 questionnaire}

Table~\ref{StatDescri1} shows some statistics for the students from France and Belgium. 

\begin{table}[h] 
\footnotesize
\caption{Comparison of a set of interesting variables between France and Belgium (2009-2010).}
\label{StatDescri1}
\begin{tabular}{lccc}
\hline\noalign{\smallskip}
Variables & France & Belgium &France+Belgium \\ 
\noalign{\smallskip}\hline\noalign{\smallskip}
Number of students & 614 & 169 & 783 \\
Success rate (\%) &  71 & 39 & 64 \\
Male/Female (\%) & 57/43  & 57/43 & 57/43 \\ 
Live with their parents (\%) & 45 & 75 & 52 \\
Hours of $\{$French, Latin, Greek$\}$ & 0.5 & 4.9 & 1.5 \\
Hours of foreign languages & 4.3 & 7.7 & 5.0 \\
Percentage of classes attended & 96 & 90 & 95 \\  
Think they will pass (\%) & 86 & 93 & 87 \\
Estimated percentage of success & 70 & 67 & 70 \\
Percentage of time devoted to studies in order to pass & 53 & 58 & 54 \\
Number of courses not attended & 0.7 & 1.2 & 0.8 \\
Never smoke (\%) &  69  & 81 & 71 \\
\noalign{\smallskip}\hline
\end{tabular}
\end{table}

We can see that of the 614 French students, 71\% will pass at the end of the academic year, compared with just 39\% in Belgium (66 out of 169 students). The male/female breakdown (57\%/43\%) is the same at both universities. Forty-five percent of the French students live with their parents, versus 75\% in Belgium. There is a large difference in terms of number of hours of $\{$French, Latin, Greek$\}$ at high school: a mean of 0.5 hours in France compared with a mean of 4.9 hours in Belgium. The same trend can be seen with the number of hours of foreign languages: 4.3 hours in France and 7.7 hours in Belgium. The French students estimate their attendance rate at 96\%, as against just 90\% for the Belgian students. Fourteen percent of the French students think they will fail, compared with only 7\% of the Belgian students (although the proportion of Belgian students who actually fail at the end of the year is higher than in France!). However, the Belgian students estimate their percentage of success at 67\%, 
versus 70\% in France. Also, in France, of the students who think they will pass (i.e. 86\%), 75\% actually do pass at the end of the year, while of those who think they will fail (i.e. 14\%), only 45\% pass at the end of year. In other words, 21\% of the French students are too optimistic, while 6\% are too pessimistic. In Belgium, of the students who think they will pass (i.e. 93\%), 41\% actually do pass at the end of the year, while of those who think they will fail (i.e. 7\%), 83\% actually do fail at the end of year. In other words, 55\% of the Belgian students are too optimistic, while only 1\% are too pessimistic.

The Belgian students think that they need to devote 58\% of their time to their studies in order to pass, compared with a lower figure of 53\% in France. The number of courses that the students do not attend (on average) is equal to 0.7 in France and 1.2 in Belgium. 

Sixty-nine percent of the French students never smoke (which means that 31\% smoke at least a little), while 81\% of the Belgian students never do. 

For other variables, not shown in Table~\ref{StatDescri1}, such as the students' results on graduating from high school, or alcohol consumption, the students in both countries were found to be very similar. 

\subsection{2010-2011 questionnaire}

We first analyze the questions in the standard questionnaire, as we did for the 2009-2010 data, and then describe the answers given to the specific questions related to student class behavior.

\subsubsection{Standard questions}

Table~\ref{StatDescri2} shows some statistics for the students from France and Belgium.

\begin{table}[h] 
\footnotesize
\caption{Comparison of a set of interesting variables between France and Belgium (2010-2011).}
\begin{tabular}{lccc}
\hline\noalign{\smallskip}
Variables & France & Belgium &France+Belgium \\ 
\noalign{\smallskip}\hline\noalign{\smallskip}
Number of students & 52 & 162 & 214 \\
Success rate (\%) &  54 & 38 & 42 \\
Male/Female (\%) & 79/21  & 65/35 & 68/32 \\
Live with their parents (\%) & 56 & 75 & 71 \\
Hours of $\{$French, Latin, Greek$\}$ & 1.2 & 5.1 & 4.2 \\
Hours of foreign languages & 3.8 & 7.2 & 6.4 \\
Percentage of classes attended & 92 & 94 & 94 \\  
Think they will pass (\%) & 75 & 97 & 92 \\
Estimated percentage of success & 66 & 68 & 67 \\
Percentage of time devoted to studies in order to pass & 50 & 57 & 55 \\
Number of courses not attended (on average) & 0.7 & 0.8 & 0.7 \\
Never smoke (\%) &  44  & 86 & 76 \\
\noalign{\smallskip}\hline
\end{tabular}
\label{StatDescri2}
\end{table}

We can see that of the 52 French students, 54\% pass. In Belgium, the rate is lower, at 38\%. At both universities, there are more men than women (79\% men in France, and 65\% in Belgium). Fifty-six percent of the French students live with their parents, versus 75\% in Belgium. There is a large difference in terms of number of hours of $\{$French, Latin, Greek$\}$ at high school: a mean of 1.2 hours in France compared with a mean of 5.1 hours in Belgium. The same trend can be seen with the number of hours of foreign languages: 3.8 hours in France and 7.2 hours in Belgium. The French students estimate their attendance rate at 92\%, similar to the figure in Belgium (94\%). Twenty-five percent of the French students think they will fail, compared with just 3\% of the Belgian students. However, the percentage of success estimated by the French and Belgian students is almost equal (66\% in France, 68\% in Belgium). Also, in France, of the students who think they will pass (i.e. 75\%), 64\% actually do, while of 
those who think they will fail (i.e. 25\%), 77\% actually do. In other words, 35\% of the French students are too optimistic, while 6\% are too pessimistic. In Belgium, of the students who think they will pass (i.e. 97\%), 38\% actually do, while of the five students who think they will fail (i.e. 3\%), only two actually do (40\%). In other words, 60\% of the Belgian students are too optimistic, while only 2\% are too pessimistic.

The Belgian students think that they need to devote 57\% of their time to their studies in order to pass, compared with 50\% in France. The number of courses that the students do not attend (on average) is similar in both countries: 0.7 in France and 0.8 in Belgium. 

Only 44\% of the French students never smoke, while in Belgium 86\% never do. 

For other variables, not shown in Table~\ref{StatDescri2}, such as the students' results on graduating from high school, hours of mathematics, or alcohol consumption, the students in both countries were found to be very similar.  

\subsubsection{Behavior}
\label{SubsubsecBehaviors}

Table~\ref{StatQuestionsBehaviors} shows the responses to the questions related to student behavior in the lecture theater, expressed as a percentage for the four possible options (i.e. ``Never'', ``Rarely'', ``Regularly'' or ``Very often''). The results are given separately for France and Belgium, with France shown first and then Belgium, separated by a slash. 

\begin{table}[h]
\footnotesize
\caption{Answers to the questions related to student behavior in the lecture theater (expressed as a percentage for the four possible options), for France and Belgium (France/Belgium).}
\begin{tabular}{lcccc}
\hline\noalign{\smallskip}
Questions & Never & Rarely & Regularly & Very often \\ 
\noalign{\smallskip}\hline\noalign{\smallskip}
Leave some lessons after the break? & 73/57 & 21/39 & 6/3 & 0/1 \\
Arrive late for classes? & 36/49 & 52/45 & 8/4 & 4/2 \\
Take notes during lectures? & 2/0 & 8/6 & 42/42 & 48/52 \\
Sit in the same area in the lecture theater? & 4/1 & 6/12 & 48/57 & 42/30 \\ 
Sit next to the same people in the lecture theater? & 4/1 & 13/4 & 35/51 & 48/44 \\ 
Bothered by the behavior of their neighbors? & 33/18 & 46/59 & 13/18 & 8/5 \\ 
Talk with their neighbors? & 13/2 & 39/58 & 38/33 & 10/7 \\
Easily distracted? & 4/7 & 44/46 & 37/36   & 15/11 \\
Ask the lecturer questions? & 40/40 & 48/52 & 2/7 & 10/1 \\ 
Ask their neighbors questions about the lesson? & 10/2 & 19/25 & 58/59 & 13/14 \\ 
Attentive from the beginning to the end of the lesson?& 6/5 & 31/23 & 48/62 & 15/10 \\ 
Use their laptop to take notes during classes? & 84/93 & 8/5 & 2/2 & 6/0 \\
Use their laptop to play during classes? & 84/92 & 8/4 & 6/3 & 2/1 \\ 
Text messaging with their cell phone during classes?& 31/16 & 29/48 & 27/28 & 13/9 \\
\noalign{\smallskip}\hline
\end{tabular}
\label{StatQuestionsBehaviors}
\end{table}

We can see that the students are generally well-disciplined (do not leave the lessons after the break, are on time for classes and take notes). They like to sit in the same place and with the same people. Of the French/Belgian students, 67\%/82\% are sometimes bothered by the behavior of their neighbors but 48\%/40\% regularly or often talk to them. About half are easily distracted. Only 12\%/8\% are used to asking the lecturer questions, but 71\%/73\% are used to asking their neighbors questions; 63\%/72\% claim to be attentive most of the time throughout the lesson. Generally, they do not use their laptops to take notes or to play during classes, but 40\%/37\% like to text their friends. 

In Table~\ref{PercSuccessQuestionsBehaviors}, we indicate the success rate, depending on the answers to the 14 questions. We have grouped the answers ``Never'' and ``Rarely'' under ``No'', and the answers ``Regularly'' and ``Very often'' under ``Yes''. For this analysis, we have grouped together the French and Belgian students to make the analysis of the results more conclusive.  

\begin{table}[h]
\footnotesize
\caption{Success rates based on the answers given to the questions related to student behavior in the lecture theater.}
\begin{tabular}{lcc}
\hline\noalign{\smallskip}
Questions & No & Yes \\ 
\noalign{\smallskip}\hline\noalign{\smallskip}
Leave some lessons after the break? & 42 & 33 \\
Arrive late for classes? & 43 &27  \\
Take notes during lectures? & 14 &44  \\
Sit in the same area in the lecture theater? & 46 & 41 \\ 
Sit next to the same people in the lecture theater? &  47& 42 \\ 
Bothered by the behavior of their neighbors? &44  & 37 \\ 
Talk with their neighbors? & 43 & 41 \\
Easily distracted? &49  & 34 \\
Ask the lecturer questions? & 40 & 61 \\ 
Ask their neighbors questions about the lesson? & 50 & 39 \\ 
Attentive from the beginning to the end of the lesson?& 36 & 45 \\ 
Use their laptop to take notes during classes? & 43 & 29 \\
Use their laptop to play during classes? & 43 & 20 \\ 
Text messaging with their cell phone during classes?& 46 & 35 \\
\noalign{\smallskip}\hline
\end{tabular}
\label{PercSuccessQuestionsBehaviors}
\end{table}

We can see that there are sometimes large differences in success rate, depending on the answers to the questions. For example, 44\% of the students who take notes during classes pass, while of those who do not take notes, only 14\% pass. Also, students who ask the lecturer questions have a 61\% chance of success, compared with 40\% for those who do not. However, these results have to be viewed with caution: those who do not take notes only represent 7\% of the students, and those who ask the lecturer questions only 8\%. \\

Concerning the position of the students in the lecture theater, Table~\ref{PercFavortivePositions} shows the nine possible positions, and the percentages of students who prefer those positions, for France and Belgium. Please note that the sum is greater than 100\%, since the students could indicate more than one favorite position. 

\begin{table}[h]
\footnotesize
\caption{Favorite positions of the students in the lecture theater (\%), for France and Belgium (France/Belgium).}
\begin{tabular}{lccc}
\hline\noalign{\smallskip}
& Left & Center & Right \\ 
\noalign{\smallskip}\hline\noalign{\smallskip}
Back & 9.6/7.4 & 5.8/21.0 & 7.7/6.2    \\
Middle & 26.9/25.9 & 65.4/67.3 & 21.2/27.2  \\
Front & 15.4/4.3 & 25.0/22.8 & 9.3/3.1 \\
\noalign{\smallskip}\hline
\end{tabular}
\label{PercFavortivePositions}
\end{table}

With a choice of seat ranging from the back to the front of the lecture theater, the students clearly prefer to sit in the middle, and the position right in the center is one of the students' favorite (65.4\%/67.3\%).

In Table~\ref{PercSuccessFavortivePositions}, we indicate the success rate according to the students' favorite positions (French and Belgian students are grouped together). We can see that the rates are much higher for the positions at the front of the lecture theater than for those at the back. 

\begin{table}[h]
\footnotesize
\caption{Success rate according to position (\%).}
\begin{tabular}{lccc}
\hline\noalign{\smallskip}
& Left & Center & Right \\ 
\noalign{\smallskip}\hline\noalign{\smallskip}
Back & 35.3 & 29.7 & 28.6   \\
Middle & 42.9 & 43.4 & 32.7 \\
Front &60.0 & 52.0 & 50.0 \\
\noalign{\smallskip}\hline
\end{tabular}
\label{PercSuccessFavortivePositions}
\end{table}

\subsection{Comparison between 2009-2010 and 2010-2011 data}

If we compare the data for 2009-2010 and 2010-2011, there are sometimes large differences. Indeed, the data is quite disparate: in 2009-2010 there were 614 French students, as against only 52 in 2010-2011. Moreover, those students came from different faculties. For example, just for the success rate, there is a marked difference: a mean of 64\% in 2009-2010, and 42\% in 2010-2011. 

On the other hand, if we compare the data for 2009-2010 and 2010-2011 only considering the Belgian students, as those students came from the same faculties, only some variables display significant differences (applying variance analysis). The variables are: the timing of the students' decision to enroll (they decided earlier in 2010-2011), the welcoming aspect of the university and the city (they find the university and the city more welcoming in 2010-2011 than in 2009-2010), the percentage of classes attended (90\% in 2009-2010, 94\% in 2010-2011), the number of courses they do not attend (1.2 in 2009-2010, 0.8 in 2010-2011), the number of courses where they think they will have problems (2.04 in 2009-2010, 1.74 in 2010-2011), and whether they think that there are enough practical courses (they are more in agreement with this in 2010-2011).

\section{Correlations}
\label{SecCorrelations}

To identify the variables that are most correlated with success, we performed the chi-squared test of independence. For each variable, the null hypothesis is that the academic success of the students does not depend on that variable. We applied this test to the different variables extracted from the questionnaire and considered a probability of 0.05 for the rejection of the null hypothesis. 

For the variables that are linked to the success of the students, we also applied the Spearman's rank correlation coefficient in order to get an order of magnitude of the relationship between the different variables and success.

\subsection{Questionnaire of 2009-2010}

In the following, we show the results for the two universities separately (France and Belgium), and then mix the students from both countries. We consider Parmentier's classification~\cite{Parmentier94} and examine three groups of variables: those related to the students' personal history, those related to the students' behavior during their studies and those related to the students' perception of academic life.  

\subsubsection{France}

Of the 145 variables, 34 were found to be linked to the success of the students at the end of the year, i.e. about one in four variables.

Six variables are related to the students' personal history: rank in their class in the final year of high school, number of hours of mathematics during the final year of high school, average grade in the final year of high school, participation in preparatory courses, their mother's qualifications and the year they graduated from high school. 

We found that four variables are associated with the students' behavior during their studies: percentage of classes attended, percentage of time they spend sleeping, whether or not they prefer to work in a group and whether they spend the weekends with their parents. 

Many variables are related to the students' perception of academic life: number of courses where they think they will have problems, whether they find the courses/teachers motivating, whether they find the lessons easy to understand, whether they think they were well prepared for university, whether they find adapting to academic life hard; some variables related to their interest in the courses, others to their own confidence (estimated percentage of success and whether they think they will pass). 

A group that includes variables concerning the way the students chose their degree program also emerged. For example, whether they chose this course of study to find a rewarding job, based on personal preference, to earn respect or because their parents will be proud of them. \\

By computing the Spearman's rank correlation coefficients between these 34 variables and success at the end of the year, we only found 7 variables that have an absolute value of coefficient $r$ greater than 0.2. The variables are: estimated percentage of success ($r$=0.28), whether they find the lessons easy to understand ($r$=0.24), rank in their class in the final year of high school ($r$=0.24), whether they think they will pass ($r$=0.23), percentage of classes attended ($r$=0.21), whether they think they made the right choice enrolling at their university ($r$=0.21), and whether they think they were well prepared for university ($r$=0.21). 

Table~\ref{SpearmanFrance} shows the difference between the students who pass ($S$=1) and those who fail ($S$=0), for the mean of the seven most correlated variables. For each of the variables, we give the set of integer values the variable can take. The higher the value, the more the students agree with the statement. For the variable ``rank in their class in the final year of high school'', a value of 5 (respectively 1) means that they were very close to being first (respectively last) in their class. 

\begin{table}[h]
\footnotesize
\caption{Difference between the students who pass ($S$=1) and those who fail ($S$=0), for the mean of the seven most correlated variables ($|r|>0.2$) (France, 2009-2010).}
\begin{tabular}{lcc}
\hline\noalign{\smallskip}
Variables & Mean ($S$=1) & Mean ($S$=0) \\
\noalign{\smallskip}\hline\noalign{\smallskip}
Estimated percentage of success $\{1,\ldots,10\}$ & 7.51 &  6.38 \\ 
Find the lessons easy to understand $\{1,\ldots,5\}$ & 3.27 & 2.74 \\ 
Rank in their class in the final year of high school $\{1,\ldots,5\}$ & 3.56 & 3.05 \\
Think they will pass $\{0,1\}$ & 0.91 & 0.73 \\ 
Percentage of classes attended $\{1,\ldots,5\}$  & 4.63 & 4.17 \\ 
Think they made the right choice enrolling here $\{0,1\}$  & 0.95 & 0.83 \\ 
Think they were well prepared for university $\{1,\ldots,5\}$  & 3.27 & 2.71 \\
\noalign{\smallskip}\hline 
\end{tabular}
\normalsize
\label{SpearmanFrance}
\end{table}

\subsubsection{Belgium}

Of the 145 variables, 14 were found to be linked to the success of the students at the end of the year, i.e. about one in ten variables. This set has seven variables in common with the set of 34 variables for France. 

Five variables are related to the students' personal history: rank in their class in the final year of high school, number of hours of mathematics during the final year of high school, average grade in the final year of high school, number of hours of $\{$French, Latin, Greek$\}$ during the final year of high school, and gender.

We found that three variables are associated with the students' behavior during their studies: percentage of classes attended, number of courses not attended and whether students try to go as fast as possible when they study. 

Five variables are related to the students' perception of academic life: whether they find the lessons easy to understand, whether they find the courses motivating, whether they think it is difficult to take notes during the lessons, whether they enjoy attending classes and whether the exams scare them.

We also found one variable linked to the way the students chose their degree program: whether they chose this course of study to deepen their knowledge in a particular area.

By computing the Spearman's rank correlation coefficients between these 14 variables and success at the end of the year, we found that nine variables have an absolute value of coefficient $r$ greater than 0.2. The variables are: rank in their class in the final year of high school ($r$=0.36), whether they find the lessons easy to understand ($r$=0.26), whether they find the courses motivating ($r$=0.26), number of hours of $\{$French, Latin, Greek$\}$ ($r$=0.26), number of hours of mathematics ($r$=0.25), percentage of classes attended ($r$=0.25), average grade in the final year of high school ($r$=0.24), whether the students try to go as fast as possible when they study ($r$=-0.22), and number of courses not attended ($r$=-0.26).

Table~\ref{SpearmanBelgium} shows the difference between the students who pass ($S$=1) and those who fail ($S$=0), for the mean of the nine most correlated variables.

\begin{table}[h]
\footnotesize
\caption{Difference between the students who pass ($S$=1) and those who fail ($S$=0), for the mean of the nine most correlated variables ($|r|>0.2$) (Belgium, 2009-2010).}
\begin{tabular}{lcc}
Variables & Mean ($S$=1) & Mean ($S$=0) \\
\noalign{\smallskip}\hline\noalign{\smallskip}
Rank in their class in the final year of high school $\{1,\ldots,5\}$ & 3.79 & 3.16 \\ 
\noalign{\smallskip}\hline\noalign{\smallskip}
Find the lessons easy to understand $\{1,\ldots,5\}$ & 3.65 & 3.18 \\ 
Find the courses motivating $\{1,\ldots,4\}$ & 3.74  & 3.35 \\ 
Number of hours of $\{$French, Latin, Greek$\}$ $\{1,\ldots,4\}$ & 3.00 & 2.85 \\ 
Number of hours of mathematics $\{1,\ldots,4\}$ & 2.56 & 2.19 \\ 
Percentage of classes attended $\{1,\ldots,5\}$ & 4.05  & 3.38 \\ 
Average grade in the final year of high school $\{1,\ldots,5\}$ & 2.71  & 2.24 \\ 
Try to go as fast as possible when studying $\{1,\ldots,4\}$ & 2.15 & 2.56  \\  
Number of courses not attended $\{1,\ldots,4\}$ & 1.85 & 2.48 \\
\noalign{\smallskip}\hline
\end{tabular}
\normalsize
\label{SpearmanBelgium}
\end{table}

\subsubsection{France + Belgium}

We also applied the chi-squared test of independence to the data for the students from both countries. The number of students in the data set is equal to 783, and the success rate is 64\%. 

Of the 145 variables, 43 were found to be linked to the success of the students at the end of the year, i.e. about three in ten variables. This set has 25 variables in common with the set composed of the 34 variables for France and the 14 for Belgium.

We shall not go into detail here, simply mentioning that all the most correlated variables for France (from Table~\ref{SpearmanFrance}) also appear among the 43 variables of the mixed set, and that seven of the nine most correlated variables for Belgium (from Table~\ref{SpearmanBelgium}) appear among the 43 variables. Only the variables ``number of hours of $\{$French, Latin, Greek$\}$'' and ``whether the students try to go as fast as possible when they study'' do not appear when considering the mixed set.

Only five of the 43 variables have a Spearman's rank correlation coefficient $r$ with an absolute value greater than 0.2: percentage of classes attended, estimated percentage of success, rank in their class in the final year of high school, participation in preparatory courses, and whether they find the lessons easy to understand.

\subsection{2010-2011 questionnaire}

In the following, we have grouped the data from France and Belgium, since there are only 52 students from France. 

Comparing with the data for 2009-2010, we also took into account the variables related to the behavior of the students during classes and the variables related to their favorite positions in the lecture theater. 

Of the 175 variables, 21 were found to be linked to the academic success of the students, i.e. about one in eight variables.

Eight variables are related to the students' personal history: rank in their class in the final year of high school, average grade in the final year of high school, number of hours of science, mathematics and social/economic sciences during the final year of high school, number of siblings that have studied at university, gender, and father's qualifications. 

We found that six variables are associated with the students' behavior during their studies: percentage of classes attended (including lectures and practical classes), percentage of lectures attended, estimated percentage of success, whether or not they prefer to work in a group, percentage of time they spend studying, and number of courses not attended. 

Two variables are related to the students' perception of academic life: number of courses where they think they will have problems at the end of the year, and whether they think that studying at university is a way to make new friends.

Three variables concerning the way the students chose their degree program also emerge: whether they chose this course of study because it is best suited to their abilities and interests, or for recognition of their degree, and whether they chose this university because its size corresponded to their aspirations. 

Two variables are related to the behavior of the students in the lecture theater: whether they talk with their neighbors, and whether they like to sit in the middle of the lecture theater. \\

By computing the Spearman's rank correlation coefficients between these 21 variables and academic success, we find that 10 variables have an absolute value of coefficient $r$ greater than 0.2. The variables are: rank in their class in the final year of high school ($r$=0.39), average grade in the final year of high school ($r$=0.31), estimated percentage of success ($r$=0.28), percentage of lectures attended ($r$=0.22), father's qualifications ($r$=0.21), number of hours of science during the final year of high school ($r$=0.21), percentage of classes attended ($r$=0.20), number of courses where they think they will have problems at the end of the year ($r$=-0.26), number of hours of social/economic sciences ($r$=-0.22), and whether or not they prefer to work in a group ($r$=-0.21).

Table~\ref{SpearmanAll} shows the difference between the students who pass ($S$=1) and those who fail ($S$=0), for the mean of the 10 most correlated variables. For each of the variables, we give the set of integer values the variables can take. The higher the value, the more the students agree with the statement. For the variable ``rank in their class in the final year of high school'', a value of 5 (respectively 1) means that they were very close to being first (respectively last) in their class. For the variable ``Father's qualifications'', the higher the value, the more qualified the father is. 

\begin{table}[h]
\footnotesize
\caption{Difference between the students who pass ($S$=1) and those who fail ($S$=0), for the mean of the 10 most correlated variables ($|r|>0.2$) (France+Belgium, 2010-2011).}
\begin{tabular}{lcc}
\hline\noalign{\smallskip}
Variables & Mean ($S$=1) & Mean ($S$=0) \\
\noalign{\smallskip}\hline\noalign{\smallskip}
Rank in their class in the final year of high school $\{1,\ldots,5\}$ & 3.84 & 3.04 \\ 
Average grade in the final year of high school $\{1,\ldots,5\}$ & 2.71 & 2.06 \\ 
Estimated percentage of success $\{1,\ldots,10\}$ & 7.39 & 6.48 \\ 
Percentage of lectures attended $\{1,\ldots,5\}$ & 4.24  & 3.65 \\ 
Father's qualifications $\{1,\ldots,4\}$ & 2.84 & 2.43 \\ 
Number of hours of science $\{1,\ldots,4\}$ & 2.38 & 1.85 \\ 
Percentage of classes attended $\{1,\ldots,5\}$ & 4.47 & 3.98 \\ 
Number of courses where they think they will have problems $\{1,\ldots,5\}$ & 2.33 & 2.76 \\ 
Number of hours of social/economic sciences $\{1,\ldots,5\}$ & 1.22 & 1.58 \\ 
Whether or not they prefer to work in a group $\{1,\ldots,5\}$  & 2.14 & 2.66 \\
\noalign{\smallskip}\hline
\end{tabular}
\normalsize
\label{SpearmanAll}
\end{table}

\subsection{Comparison between 2009-2010 and 2010-2011}

We compared the variables that are most correlated with success between the academic years 2009-2010 and 2010-2010, and found that about one in three variables for 2009-2010 also appear in 2010-2011. Two variables are highly correlated with success in both data sets: rank in their class in the final year of high school and average grade in the final year of high school. 

We also compared the variables that are most correlated with success between the Belgian students in the two academic years studied, but again found that only about one in three variables for 2009-2010 also appear in 2010-2011, even though those students come from the same faculties at the same university.

\section{Methods}
\label{SecMethods}

Six traditional supervised data mining methods were considered to predict academic success according to the different variables extracted from the questionnaires. All these methods were implemented with the \textsf{R} free software environment for statistical computing~\cite{Torgo10}. In the following, we briefly describe each of the data mining methods and how they were implemented.  

\subsection{Decision trees}

A decision tree~\cite{Breiman84} is a hierarchy of logical tests on some of the predictor variables. A tree starts with a \emph{root node}, which is split into two new nodes. Each node of a tree has two branches, except the \emph{leaf nodes}, which are terminal nodes of the tree. The branches are related to the outcome of a test on one of the predictor variables. Each leaf node represents a value of the decision variable. Consequently, to predict the value of the decision variable in a specific case, it is enough to follow the tree according to the predictor variables and the rules of the branches. Once a leaf node is reached, the prediction associated with the case in question is obtained. 

Decision trees have the ability to create a model that totally corresponds to the training set, i.e. with a resubstitution error equal to zero. However, such a model generally has a high validation error, due to what is called \emph{overfitting}. It is often better to use a subtree, which yields better predictions in general and for new cases.  

We used the \textsf{rpart} package to implement decision trees in \textsf{R}. The decision trees built with this package use \emph{cost complexity} pruning~\cite{Breiman84} in order to achieve the best compromise between predictive accuracy and tree size. 

In our study we considered two variants of the pruning rule; for each data set, two different decision trees will therefore be developed.  

\subsection{Discriminant analysis}

The linear discriminant analysis model~\cite{Friedman89} tries to find a linear combination of features that characterize each case to be classified. The resulting combination is then used as a linear classifier. 

We used both linear discriminant analysis and quadratic discriminant analysis through the functions \texttt{lda} and \texttt{qda} of the package \textsc{\textsf{mass}} of \textsf{R}.  %

\subsection{Random forests}

Random forests~\cite{Breiman01} consist of a set of decision trees. The number of trees in the set is a parameter of the method. For each tree, a different sample of data from the original data set is selected, to form training sets. A tree is then built from this training set. For each node of the trees, only a random subset of the predictors is chosen. When all the trees have been built, the predictions are obtained by averaging out the predictions of each tree. For classification problems, this is similar to a voting mechanism (the class that gets more votes across all the trees is the prediction). 

We used the \textsf{randomForest} package to implement random forests in \textsf{R}. We will build forests of 1000 trees. The number of predictors considered at each node was optimized, with a simple enumeration algorithm, in order to obtain the best value for the resubstitution error and the best value for the cross-validation error (the optimum number of predictors can be different). 

\subsection{Logistic regression}

Logistic regression~\cite{Hilbe09} is similar to multiple linear regression, but the fact that the dependent variable is categorical is taken into account. Eventually, for each case of the data set, a probability is associated with the decision variable. In our case, as the decision variable is binary, if the probability is higher than 0.5, we consider that the student will pass, and otherwise fail.

We used the function \texttt{glm} of \textsf{R} to apply logistic regression to our data. 

\subsection{Support vector machines}

Support vector machines~\cite{Vapnik98} use an implicit mapping of the input data into a high-dimensional feature space defined by a kernel function. Once the mapping is done, a simple linear method is applied to the data, but in this high-dimensional feature space non-linearly related to the input space.

In \textsf{R}, we used the \textsf{kernel} package~\cite{Karatzoglou04}, with a Gaussian radial-based kernel function. We also used the support vector machine predictor obtained with the package \textsf{e1071}, which provides an interface to \textsf{libsvm}, a library containing the most popular support vector machine formulations. 

\subsection{$k$-nearest neighbors}

The $k$-nearest neighbors technique is quite particular, since no model is built from the training data ($k$-nearest neighbors belong to the class of \emph{lazy learner} methods). The main work of this method happens at prediction time. Given a new test case, its prediction is obtained by searching for similar cases in the training data stored. The $k$ most similar cases are used to obtain the prediction for the given test case. The Euclidean distance is often used to measure the similarity between the cases. 

To implement this method, we used the function \texttt{knn()} of the package \textsf{class}~\cite{Venables02} of \textsf{R}. The parameter $k$ was optimized, with a simple enumeration algorithm, in order to obtain the best value for the cross-validation error. 

\section{Results}

\label{SecResults}

In this section, we explain the results of the different data mining methods used in order to predict student academic success. Our aim is mainly to obtain a model that performs well in practice and can achieve high classification rates for new data sets, and not only for the data set on which the model is based (measured by the resubstitution error). 

In order to estimate the power of classification of the models in practice, we use $k$-fold cross-validation~\cite{Refaeilzadeh09}. This method involves separating the data into two groups: a small group (equal in size to the total number of students divided by $k$), called the test, or validation set, and a large group (equal in size to the total number of students multiplied by $(\frac{k-1}{k})$), called the training set. The training set is used to construct the model, while the validation set is used to evaluate the model. This process is repeated $k$ times. We used $k$=10, since it is a value that gives good results, as shown by empirical studies~\cite{Kohavi95}. Stratification was also considered: we ensured that the proportion of success is the same in both the training and the validation sets.

\subsection{2009-2010 questionnaire}

We first analyze the 2009-2010 questionnaire, and compare the results obtained in France and Belgium. 

\subsubsection{Selection of the variables}

\label{SubsubSectionSelection}

We used the following rule to select the variables that participate in the creation of the models. For each data set (France, Belgium, and France+Belgium), we retained the variables that have a link with success according to the chi-squared test of independence, with $p=0.05$. We also considered the variables that have a Spearman's rank correlation coefficient greater, in terms of absolute value, than 0.15. 

Using this rule, for the data set for France, 34 variables were used; for Belgium, 23 variables; and for the mixed set composed of the data from France and Belgium, 43 variables.  

\subsubsection{Results in resubstitution}

In Table~\ref{ResubstitutionError0910}, we show the resubstitution error obtained using the different data mining methods. 

\begin{table}[h]
\small
\caption{Resubstitution error ($\%$) for the different data mining methods (2009-2010).}
\begin{tabular}{lccc}
\hline\noalign{\smallskip}
Method & France & Belgium & France + Belgium \\
\noalign{\smallskip}\hline\noalign{\smallskip}
Decision Tree 1 &  19.22& 19.53 & 20.95 \\
Decision Tree 2 & 29.32 & 18.34 & 24.39 \\
Linear Discriminant Analysis & 22.99 & 19.53 &25.93 \\
Quadratic Discriminant Analysis & \textbf{15.31} & \textbf{5.33} & 14.43 \\
Random Forests & 22.48 & 21.89 &24.39 \\
Logistic Regression & 22.31 & 16.57 &23.37 \\
Support Vector Machine 1 & 16.94 & 11.83 & \textbf{13.67} \\
Support Vector Machine 2 & 22.80 & 20.12 & 21.46  \\
\noalign{\smallskip}\hline
\end{tabular}
\normalsize
\label{ResubstitutionError0910}
\end{table}

For the French data set and the Belgian data set, the best results are obtained using quadratic discriminant analysis, with a resubstitution error of 15.31\% and 5.33\%, i.e. a correct classification rate of 84.69\% and 94.67\%. For the set composed of the data from France and Belgium, the best resubstitution error is obtained using the first version of the support vector machine method, with an error equal to 13.67\%, i.e. a correct classification rate of 86.33\%. 

The correct classification rates are quite good, all greater than 84\%, but what is quite disappointing is that the resubstitution error of the set for France and Belgium is always higher than either the resubstitution error of the Belgian set or the resubstitution error of the French set. This means that working with a data set of higher size does not help to improve the resubstitution error of either the French or the Belgian set. 

For example, for decision tree 2, if we take the set for France and Belgium, we have a total error of 24.39\%. If we use the tree produced and compute the corresponding error for the French set and for the Belgian set, we obtain errors equal to 23.45\% and 27.81\% respectively. We have improved the results for France (going from 29.32\% to 23.45\%), but we have worsened the results for Belgium (18.34\% to 27.81\%). 

For decision tree 1, it is even worse, since both errors have deteriorated (going from 19.22\% to 20.20\% for France, and from 19.53\% to 23.67\% for Belgium). 

\subsubsection{Results in cross-validation}

Table~\ref{CrossValidationError0910} shows the cross-validation errors (mean and standard deviation) for the different data mining methods. For the French set, the best results are obtained with the random forests method (correct classification rate equal to 77.7\%), and for the Belgian set, with the support vector machine method (version 1, correct classification rate equal to 78.18\%). For the French and Belgian set, the best results are obtained using the random forests method, with a correct classification rate equal to 75.9\%. Once again, however, the cross-validation error of the French and Belgian set is always greater than either the cross-validation error of the Belgian set or the cross-validation error of the French set. 

\begin{table}[h]
\small
\caption{Cross-validation error ($\%$) for the different data mining methods (2009-2010).}
\begin{tabular}{lccc}
\hline\noalign{\smallskip}
Method & France & Belgium & France + Belgium  \\
\noalign{\smallskip}\hline\noalign{\smallskip}
Decision Tree 1 &  29.84 $\pm$ 3.92 & 28.48 $\pm$ 9.24  & 32.44 $\pm$ 3.58 \\
Decision Tree 2 & 30.00 $\pm$ 1.56 & 33.33 $\pm$ 7.73 & 33.72 $\pm$ 4.68  \\
Linear Discriminant Analysis & 25.74 $\pm$ 4.64 & 24.85 $\pm$ 3.95 & 25.64 $\pm$ 4.80 \\
Quadratic Discriminant Analysis & 29.02 $\pm$ 7.00 & 33.33 $\pm$ 3.71 & 32.18 $\pm$ 4.16 \\
Random Forests & \textbf{22.30 $\pm$ 3.80} &  22.42 $\pm$ 3.46 & \textbf{24.10 $\pm$ 2.89} \\  
Logistic Regression & 27.05 $\pm$ 4.25 &  23.03 $\pm$ 5.91 &  26.28 $\pm$ 4.96 \\
Support Vector Machine 1 & 25.08 $\pm$ 3.37 & \textbf{21.82 $\pm$ 9.44} & 25.26 $\pm$ 4.84 \\
Support Vector Machine 2 & 26.39 $\pm$ 3.90 & 22.42 $\pm$ 5.07 & 24.87 $\pm$ 4.65 \\  
$k$-Nearest Neighbors & 26.64 $\pm$ 2.38 &24.44  $\pm$ 3.52 &  29.23  $\pm$ 4.05 \\
\noalign{\smallskip}\hline
\end{tabular}
\normalsize
\label{CrossValidationError0910}
\end{table}

\subsection{2010-2011 questionnaire}

For the 2010-2011 questionnaire, we grouped the data for France and Belgium, since the data set for France is composed of only 52 students. The aim of this section is to see whether the variables related to student behavior can improve the classification rates. 

\subsubsection{Selection of the variables}
\label{SecVariables}

Three sets of variables were considered: one with all the variables (called \texttt{\{All\}} set), one with all the variables except those related to student class behavior (called \texttt{\{All\} $\backslash$ \{Behavior\}} set), and one with only the variables related to student class behavior (called \texttt{\{Behavior\}} set). 

For the first two sets, since their cardinality is rather important, we used the same rule as in Section~\ref{SubsubSectionSelection} to reduce the size of both sets. For the first set, 35 variables were selected, and 31 for the second. The four variables related to behavior are: whether students talk with their neighbors, whether they like to sit in the middle of the lecture theater, whether they are easily distracted, and whether they take notes during lectures. 

The \texttt{\{Behavior\}} set contains 29 variables (i.e. the 14 variables related to the 14 questions listed in Table~\ref{StatQuestionsBehaviors} and the 15 variables related to the position of the students in the lecture theater).

\subsubsection{Results in resubstitution}

In Table~\ref{ResubstitutionError1011}, we show the resubstitution error obtained with the different data mining methods, and for the three subsets of variables considered. 

\begin{table}[h]
\small
\caption{Resubstitution error ($\%$) for the different data mining methods (2010-2011).}
\begin{tabular}{lccc}
\hline\noalign{\smallskip}
Method & \texttt{\{All\}} & \texttt{\{All\} $\backslash$ \{Behavior\}} & \texttt{\{Behavior\}} \\
\noalign{\smallskip}\hline\noalign{\smallskip}
Decision Tree 1 & 18.22 &  18.22&  24.77   \\   
Decision Tree 2 &30.37  &  30.37   & 42.06 \\
Linear Discriminant Analysis &19.63  & 19.63   & 27.57 \\
Quadratic Discriminant Analysis & \textbf{4.21}  & \textbf{4.21}  & \textbf{16.82} \\
Random Forests &23.83  & 25.70  &  35.51 \\ 
Logistic Regression &  18.22& 19.16   & 28.50 \\
Support Vector Machine 1 &13.55  &  13.55   & 20.09  \\
Support Vector Machine 2 &21.03  & 21.96  & 30.37 \\
\noalign{\smallskip}\hline
\end{tabular}
\normalsize
\label{ResubstitutionError1011}
\end{table}

For all the sets, the best results are obtained using quadratic discriminant analysis, with a resubstitution error of 4.21\%, 4.21\% and 16.82\% respectively. Furthermore, for all the methods, we can see that the results with the \texttt{\{All\}} set are better than those with the \texttt{\{All\} $\backslash$ \{Behavior\}} set, and the results with the \texttt{\{All\} $\backslash$ \{Behavior\}} set are better than those with the \texttt{\{Behavior\}} set. 

\subsubsection{Results in cross-validation}

\begin{table}[h]
\small
\caption{Cross-validation error ($\%$) for the different data mining methods (2010-2011).}
\begin{tabular}{lccc}
\hline\noalign{\smallskip}
Method & \texttt{\{All\}} & \texttt{\{All\} $\backslash$ \{Behavior\}} & \texttt{\{Behavior\}}  \\
\noalign{\smallskip}\hline\noalign{\smallskip}
Decision Tree 1 & 40.00 $\pm$ 10.09 & 40.00 $\pm$ 10.09  & 42.86 $\pm$ 8.09 \\
Decision Tree 2 & 30.48 $\pm$ 10.58 & 30.48 $\pm$ 10.58   & 43.81 $\pm$ 2.01  \\
Linear Discriminant Analysis & 32.86 $\pm$  9.90&  30.48 $\pm$ 8.75  &   37.14 $\pm$ 10.24 \\
Quadratic Discriminant Analysis & 40.00 $\pm$ 5.12   &  36.67 $\pm$ 6.37  & 40.00 $\pm$ 7.84 \\
Random Forests &  \textbf{24.29 $\pm$ 6.90} &   \textbf{24.76 $\pm$ 8.34}  & \textbf{34.29 $\pm$ 9.99} \\ 
Logistic Regression &  33.33 $\pm$ 10.53 & 31.90 $\pm$ 9.80 & 37.14 $\pm$ 10.48  \\
Support Vector Machine 1 & 29.52 $\pm$ 5.85  & 28.10 $\pm$ 6.90 & 37.14 $\pm$ 9.73 \\
Support Vector Machine 2 & 29.52 $\pm$ 11.40  & 29.05 $\pm$ 12.18 & 38.57 $\pm$ 11.10 \\  
$k$-Nearest Neighbors &  27.86 $\pm$ 10.21 &  27.72 $\pm$ 10.64   & 42.38 $\pm$ 11.76  \\
\noalign{\smallskip}\hline
\end{tabular}
\normalsize
\label{CrossValidationError1011}
\end{table}

Table~\ref{CrossValidationError1011} shows the cross-validation errors (mean and standard deviation) for the different data mining methods, and for the three subsets of variables considered. 

For the three sets, \texttt{\{All\}}, \texttt{\{All\} $\backslash$ \{Behavior\}} and \texttt{\{Behavior\}}, the best results are obtained using the random forests method, with a correct classification rate of 75.71\%, 75.24\% and 65.71\% respectively.

We can also see that there is practically no difference between the results obtained with the \texttt{\{All\}} and \texttt{\{All\} $\backslash$ \{Behavior\}} sets. 

The results obtained with the \texttt{Behavior} set are quite poor, since a simple and unsophisticated method of predicting the failure of each student would achieve a correct classification rate of 58.41\%, which is only slightly lower than the results obtained with the \texttt{\{Behavior\}} set and the random forests method (correct classification rate of 65.71\%).

\subsection{Data sets for 2009-2010 and 2010-2011}

In this section, we combine the data from 2009-2010 and 2010-2011 into just one set. We then try to predict the academic success of the 997 students in this set. We used the same rule as in Section~\ref{SubsubSectionSelection} to select the variables that appear in the predictions. The behavior variables from the 2010-2011 set have not been considered since they have not been integrated into the 2009-2010 data set. 

We give the results first in resubstitution and then in cross-validation. 

\subsubsection{Results in resubstitution}

In Table~\ref{ResubstitutionErrorAll}, we show the resubstitution error obtained with the different data mining methods. We compare the results obtained with the mixed set, integrating all the data (2009-2011 set) with the 2009-2010 and 2010-2011 sets.  

\begin{table}[h]
\small
\caption{Resubstitution error ($\%$) for the different data mining methods (2009-2011).}
\begin{tabular}{lccc}
\hline\noalign{\smallskip}
Method & 2009-2010 & 2010-2011 & 2009-2011 \\
\noalign{\smallskip}\hline\noalign{\smallskip}
Decision Tree 1 & 20.95 &18.22  & 27.48  \\   
Decision Tree 2 &24.39  & 30.37 & 28.59 \\
Linear Discriminant Analysis & 25.93 & 19.63 &26.78 \\
Quadratic Discriminant Analysis & 14.43 & \textbf{4.21} & \textbf{13.54} \\
Random Forests & 24.39 & 25.70 & 24.27 \\ 
Logistic Regression &23.37  & 19.16   & 22.57 \\
Support Vector Machine 1 & \textbf{13.67}  & 13.55   & 14.84 \\
Support Vector Machine 2 &21.46  & 21.96 & 21.87 \\
\noalign{\smallskip}\hline
\end{tabular}
\normalsize
\label{ResubstitutionErrorAll}
\end{table}

For the mixed set, the best results are obtained using quadratic discriminant analysis, with a correct classification rate equal to 86.46\%. We can see that only with the random forests method is the resubstitution error of the mixed set lower than the resubstitution error of both data sets for 2009-2010 and 2010-2011. 

\subsubsection{Results in cross-validation}

The cross-validation errors obtained with the different data mining methods are shown in Table~\ref{CrossValidationErrorAll}. 

\begin{table}[h]
\small
\caption{Cross-validation error ($\%$) for the different data mining methods (2009-2011).}
\begin{tabular}{lccc}
\hline\noalign{\smallskip}
Method & 2009-2010 & 2010-2011 & 2009-2011 \\
\noalign{\smallskip}\hline\noalign{\smallskip}
Decision Tree 1 & 32.44 $\pm$ 3.58  &  40.00 $\pm$ 10.09  &  30.10  $\pm$ 3.80 \\
Decision Tree 2 &  33.72 $\pm$ 4.68  &  30.48 $\pm$ 10.58   & 29.80 $\pm$ 3.60 \\
Linear Discriminant Analysis &  25.64 $\pm$ 4.80  & 30.48 $\pm$ 8.75  & 25.96  $\pm$ 2.18 \\
Quadratic Discriminant Analysis & 32.18 $\pm$ 4.16   &  36.67 $\pm$ 6.37  & 29.29 $\pm$ 3.69  \\
Random Forests &  \textbf{24.10 $\pm$ 2.89}  &   \textbf{24.76 $\pm$ 8.34}   & \textbf{23.43 $\pm$ 3.92} \\ 
Logistic Regression &  26.28 $\pm$ 4.96 &  31.90 $\pm$ 9.80  & 26.06 $\pm$ 2.12 \\
Support Vector Machine 1 & 25.26 $\pm$ 4.84   &  28.10 $\pm$ 6.90 &  24.85 $\pm$ 4.81  \\
Support Vector Machine 2 &  24.87 $\pm$ 4.65   & 29.05 $\pm$ 12.18  & 26.36 $\pm$ 2.75 \\  
$k$-Nearest Neighbors &  29.23  $\pm$ 4.05 &  27.72 $\pm$ 10.64    & 31.54  $\pm$ 3.51 \\
\noalign{\smallskip}\hline 
\end{tabular}
\normalsize
\label{CrossValidationErrorAll}
\end{table}

The best results are obtained using the random forests method, with a correct classification rate equal to 76.57\%. 

For six methods out of nine, the results obtained with the mixed set are better than the results of both data sets for 2009-2010 and 2010-2011, which is quite encouraging. Therefore, for these methods, adding the results of an additional year helps to improve the classification rates in cross-validation.

\section{Conclusion}

\label{SecConclusion}

In this study, we compared the results of data mining methods to predict the academic success of students from Belgium and France. We used two data sets from two academic years: one from 2009-2010 and the other from 2010-2011. 

In cross-validation, the best results are obtained in most cases using the random forests method, with a correct classification rate equal to around 75\%. 

We saw that considering a set composed of students from France and Belgium does not help to improve the results of the sets comprising only those students from France or Belgium. 
On the other hand, integrating the results of an additional year is effective.

We also studied the integration of variables related to the behavior of the students during classes, and their favorite positions in the lecture theater. 
We showed that correlations exist between those variables and academic success, but are only supported by some specific students and are not really helpful in predicting academic success using data mining methods. 

Thanks to this study, we are now able to create a smaller questionnaire, composed only of questions whose answers are strongly linked with success or failure. This smaller questionnaire would be easier for the students to complete and we could hope to get more accurate answers.

We also plan to use the prediction models obtained in this study to implement a decision-making tool in order to predict the students' success in the following academic year and to be able to help, right from the beginning of the year, those students identified as being at high risk of failing.

\end{document}